# Solstice and Solar Position observations in Australian Aboriginal and Torres Strait Islander traditions

Duane W. Hamacher [1,2], Robert S. Fuller [3], Trevor M. Leaman [3], and David Bosun [4]

[1] School of Physics, University of Melbourne, Parkville, VIC 3010, Australia
[2] ARC Centre of Excellence in All Sky Astrophysics in Three-Dimensions (ASTRO 3D)
[3] School of Humanities and Languages, University of New South Wales, Sydney, NSW 2052, Australia
[4] Councillor, Kubin Village, Moa, QLD, 4875, Australia

**Corresponding author**: duane.hamacher@unimelb.edu.au

**Abstract:** A major focus of the archaeoastronomical research conducted around the world focuses on understanding how ancient cultures observed sunrise and sunset points along the horizon, particularly at the solstices and equinoxes. Scholars argue that observations of these solar points are useful for developing calendars, informing ritual/ceremonial practices, and predicting seasonal change. This is the foundation of the Eurocentric four-season Julian (and later Gregorian) calendar. Famous examples include Stonehenge, Newgrange, Chichen Itza, and Chankillo. Studies at these and other sites tend to focus on solar point observations through alignments in stone arrangements, and the orientations of monuments. Despite the ongoing study of Indigenous Knowledge in Australia revealing a wealth of information about Aboriginal and Torres Strait Islander observations and interpretations of solar, lunar, and stellar properties and motions, very little has been published about the importance and use of solar point observations. The authors examine this topic through four case studies, based on methodological frameworks and approaches in ethnography, ethnohistory, archaeology, and statistics. Our findings show that Aboriginal and Torres Strait Islander people observe the solstices and other significant sunrise/sunset points along the horizon for timekeeping and indicating seasonal change – but in ways that are rather different to the four-season model developed in Western Europe.

## 1  Introduction

For the last decade, researchers have been working closely with Aboriginal and Torres Strait Islander elders to learn more about their astronomical knowledges and traditions. This has revealed a wealth of information on the subject, with a focus on the scientific and social information encoded within them (Hamacher 2012; Norris 2016). This includes knowledge about the complex motions of the Sun, Moon, stars, and planets (e.g. Hamacher and Norris, 2011; Hamacher 2015; Hamacher and Banks 2018, respectively). This research is highly multidisciplinary, drawing largely from archaeology, ethnography, history, and Indigenous studies. These disciplines serve as the methodological and theoretical foundation of cultural astronomy, which itself is generally divided between the sub-disciplines of *archaeoastronomy* (astronomical knowledge of ancient cultures and societies) and *ethnoastronomy* (astronomical knowledge of contemporary cultures and societies).

The academic discipline of archaeoastronomy developed to study how cultures and societies in the past conceptualised and utilised the motions of astronomical objects and phenomena for practical and social purposes (Ruggles 2011), focusing primarily on the archaeological record. One of the goals of archaeoastronomy is to better understand how ancient cultures developed calendars based on the movements of the Sun, Moon, and stars. The modern Western four-season calendar is based on the movements of the Sun throughout the year: the solstices mark the start of summer and winter, and the equinoxes mark the start of spring and autumn. Archaeoastronomical research seeks to better understand if ancient cultures observed these solar points and, if so, examine how they conducted these observations and made use of this knowledge.

Studies from around the world reveal that many ancient and Indigenous cultures did (and continue to) observe sunrise and sunset points for calendric purposes (Ruggles 2005). Research shows that the tracking of solar positions throughout the year primarily falls into two general categories: the first is to utilise the landscape, natural features, or existing *in situ* objects to mark out significant solar points. For example, a community may choose a place from which an observer can see the sunrise or sunset at solstices and equinoxes with respect to geographical features, such as peaks and dips in the profile of the horizon. Famous examples include sites in Bronze Age Scotland (Higginbottom et al. 2015). This requires close observations of sunrise/set azimuths throughout the year and careful selection of a location where the solar points are distinctly observable. The second is to construct a human-made feature, such as a monument, tomb, or stone arrangement for tracking solar positions across the horizon. Famous examples include Newgrange, Stonehenge, and Chankillo (Ruggles 1999; Aveni 1997; Ghezzi and Ruggles 2007). A third example is modifying a natural feature for the same purpose.

The most rigorous evidence for ancient and Indigenous observations of solar points comes directly from the source in the form of ethnographic and historical records. This enables us to know the meaning and purpose of these observations from those who built them. In their absence, we must rely on the material and archaeological record. This was the critical distinction that lead to archaeoastronomy splitting into two distinct fields. It was during the world's first dedicated conference on the subject at Oxford University in 1981 that these two methodological approaches distinguished themselves. The different approaches were primarily driven by geography, with studies in the Americas being more focused on combining archaeology with history and ethnography, while and studies in Africa and Eurasia were focused on statistical probabilities of celestial alignments in stone arrangements and other archaeological sites.

Publication of the conference's proceedings solidified the methodological divide in colour-coded fashion: *Archaeoastronomy in the New World* (Aveni, 1982) was published in a brown-covered volume, while *Archaeoastronomy in the Old World* (Heggie, 1982) was published in a green-covered volume. Hence, the fields became known as *Brown Archaeoastronomy* and *Green Archaeoastronomy*, where the former was more anthropologically/ethnographically focused and the latter was more archaeologically focused. Today, these methodological approaches are generally referred to as *ethnoastronomy* and *archaeoastronomy*, respectively, falling under the collective banner of *cultural astronomy*.

An example of Green Archaeoastronomy relates to prehistoric Britain. Few, if any, historic records exist about ancient Britons. Therefore, studies about their possible observations of solstice and equinox points focuses on the archaeological record. If the people utilised natural features such as the local landscape to mark solar points, demonstrating this generally relies heavily on statistical methodologies (Higginbottom and Clay 2016). If a statistically significant number of monumental or stone arrangements are found to cluster around a particular azimuth range, such as near the solar points, then one could present a convincing case that the culture(s) in question did so deliberately. Similarly, the engineers may have chosen a location where solar points are noted by significant horizon landscape features, such as peaks or dips – or some combination of natural and human-made features. Here, statistical models could be used to overcome coincidences and chance alignments by searching for clusters in larger datasets.

In "one off" alignment cases, the probability of the alignment being deliberate rather than chance could be considered in light of its complexity. For example, two stones standing in a field that roughly align to the summer solstice sunrise and winter solstice sunset were plausibly constructed for this purpose, but the alignment could be coincidental. This is not a strong case for deliberate solar orientations. However, some tombs and ritual passageways were built so that light from the Sun illuminates a specific chamber or motif only at solstice sunrise, such as Newgrange.

In northern Peru, the Chankillo complex consists of 13 regularly spaced, human-made towers that extend along a ridgeline with numerous surrounding buildings and structures. Constructed in the 4$^{th}$ century BCE, the towers denote the setting sun throughout the year, with the range of the towers marking out the range of the setting Sun throughout the year. Ghezzi and Ruggles (2007) demonstrated that the solar alignments were deliberate and identified a clear observing platform from which the setting Sun was seen throughout the year. In both cases, the precision needed to accomplish the feat suggests the probability of the alignments being coincidental is null.

## 2    Australian Case Studies

Is there any ethnographic, historical, or archaeological evidence that Aboriginal and Torres Strait Islander people observed and made use of the solstices and equinoxes? If so, do they utilise the surrounding landscape or did they construct a monument or stone arrangement to mark them? There is every reason to suppose that Indigenous peoples observed and noted that the Sun rose and set at different azimuths throughout the year, reaching extreme northern and southerly points before moving back across the horizon again. Despite this, very little research has been published about observations of solar rise/set points by Indigenous Australians. Recent archaeological, historic, and ethnographic research reveals that Aboriginal and Torres Strait Islander communities did indeed observe solar points and applied this to predicting seasonal change.

This paper explores four case studies, in descending order of "certainty" regarding Indigenous Knowledge about solar point observations, drawing from current ethnographic fieldwork to statistical correlations of cultural sites. The first case study involves recent ethnographic fieldwork with a Torres Strait Islander custodian and co-author on this paper (Bosun) to examine how Torres Strait Islanders observe sunset positions from the village of Kubin on Mua Island. The second case study examines ethnohistoric records of Aboriginal people in northern New South Wales observing the solstices. The third case study discusses how an archaeoastronomical survey and statistical analysis of the Wurdi Youang stone arrangement in Victoria indicates it was constructed to mark the positions of the setting Sun at the solstices and equinoxes. The final case study involves a developing statistical methodology to examine correlations between Wiradjuri cultural sites in central New South Wales and the position of the rising and setting Sun along the horizon profile.

### 2.1    Case Study 1: Mua Island, Torres Strait, Queensland

The first author (Hamacher) has been working with communities in the Torres Strait since 2014 to learn about their astronomical knowledge. During this time, elders have revealed a significant amount of traditional knowledge, particularly about the Sun. In December 2017 and July 2018, Hamacher interviewed Mua councillor and co-author David Bosun. Bosun explained how people in the village of Kubin on the southwestern shore of Mua in the western Torres Strait observe the position of the setting Sun throughout the year to inform seasonal change and food economics. He explained that Kubin villagers observe where the Sun sets with respect to the archipelago of islands to the west and southwest (Fig. 1).

Additional information is sourced from the notebooks of Eseli (1998). Peter Eseli (1886-1958), a Mabuyag man and son of Peter Papi, was one of the three chief assistants to the A.C. Haddon expedition sponsored by Cambridge University in the late 19th century. Eseli's notebooks, which were translated into English from the Kala Lagow Ya language, provide a wealth of knowledge about traditional seasonal knowledge in the western Torres Strait. By merging Bosun's descriptions with details of seasons described by Eseli as seen from Kubin, we can gain a more complete picture of how the changing positions of the setting Sun are utilised by Kubin villagers in the past and present.

Where the Sun sets along the horizon throughout the year, as well as its daily direction (i.e. moving North or South), is utilised by Kubin villagers to mark seasonal change, which also informs weather patterns, animal behaviour, and agriculture. According to Bosun, villagers note the northern and southern most setting position at the solstices. However, it is not necessarily the solsticial or equinoctial points that are important. Rather, it is the azimuths of the setting Sun relative to the islands, and whether the Sun is setting at increasingly northward or southward azimuths each day, that signals changes in seasons.

The azimuth (Az) of the setting Sun as seen from Kubin ranges from 293.7° at winter solstice (northernmost point) to 246.2° at summer solstice (southernmost point), with a full range ($\Delta$Az) of 47.5° (Fig. 2). As described by Bosun, the Sun sets over Tuin (Barney Island) at the winter solstice, and along the southern end of Zurath (Phipps Island) at the summer solstice. Although Mua Islanders note the extreme setting positions of the Sun at the solstices, the meaning behind this not the same as seasonal markers used in Western traditions. At at the winter solstice the Sun sets behind a hill on Tuin (Barney

Island). As the Sun gradually moves south, the people know the cooler southeasterly trade winds will begin shifting to the duldrum (a period marked by hot, still air) as the Sun sets between the islands of Tuin and Matu. This occurs from September (sunset Az ≈ 278°) to October (Az ≈ 258°).

The setting Sun gradually moves at southward azimuths each day. When it sets over Matu, it signals the start of the turtle mating season (*Soewlal*). This takes place from mid-October (Az ≈ 261°) to late November (Az ≈ 248°), coinciding with azimuths between Matu and Zurat. This is between the end of *Wooewra* and the start of the *Kuki*, when mating turtles float on the surface of the water and are easily caught and speared (Eseli 1998:19). The Sun reaches its southern most azimuth (Az = 246.2°) on the summer solstice, setting along the southern tip of Zurath, where it appears to meet with Kulbai Kulbai (Spencer Island). This signals that the wet monsoon season is at hand (which begins in late December).

At this point, the setting Sun begins shifting further to the north each night. Eseli (1998) notes that *Wooewra*, the southeast trade wind (also called *Sager*), blows during the drier months of March (Az ≈ 262°) through November (Az ≈ 255°). During this time, the food supply is more abundant and travel is easier. Sunset in March occurs just north of Matu. The time for harvest is called *Kek*, that heralds the appearance of new yams and occurs from March to April (Az ≈ 275°), between the *Kuki* and *Wooewra*. It is associated with the heliacal rising of the star Achernar (Alpha Eridani) in mid-April:

> *"When the rising of a star is expected it is the duty of the old men to watch. They get up when the birds begin to cry and watch till daybreak."*
> - W.H.R. Rivers in Haddon (1912: 224).

At this time, the sun sets along the southern edge of Tuin and Ngul. The setting Sun gradually slows until it reaches its northern most azimuth at the winter Solstice in late June. At this time, the Sun again sets over the northern half of Tuin.

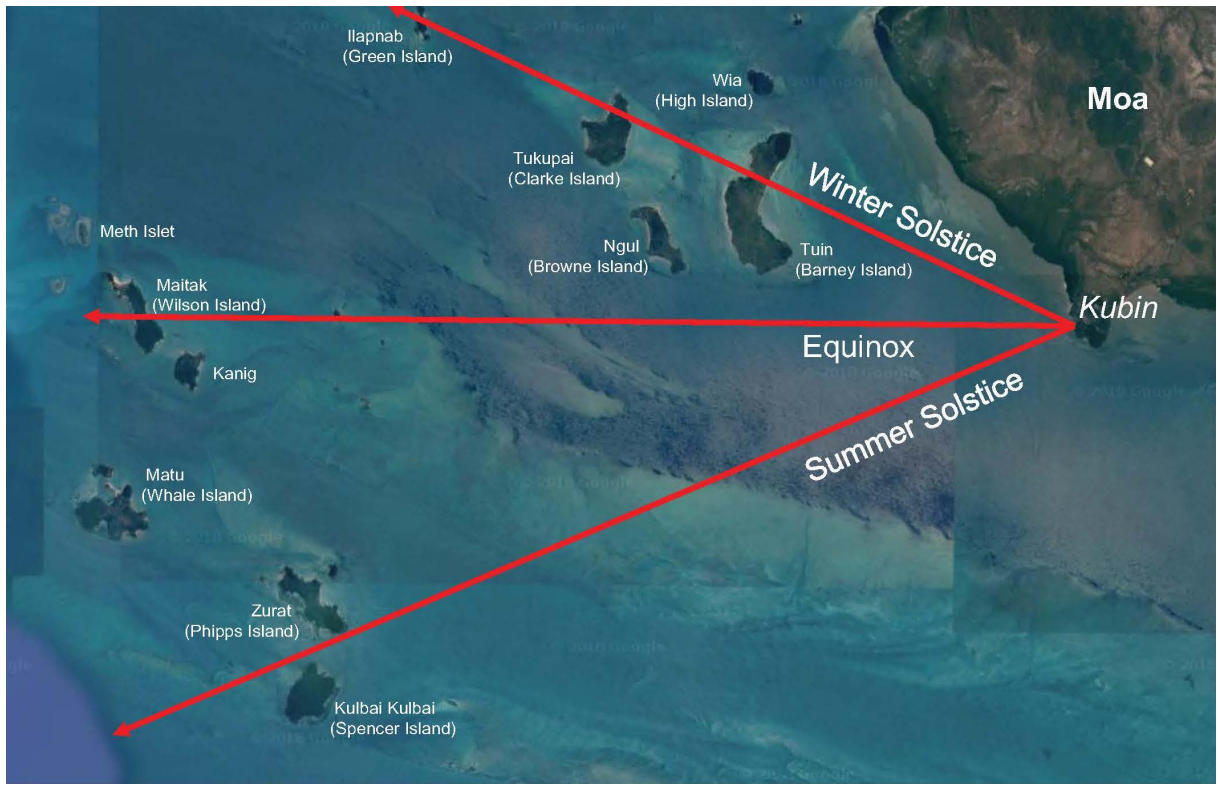

**Fig. 1:** Islands to the west and southwest of Mua Island, western Torres Strait featuring their Western and traditional names (where available), with the solstice and equinox lines in red. Image modified from Google Earth.

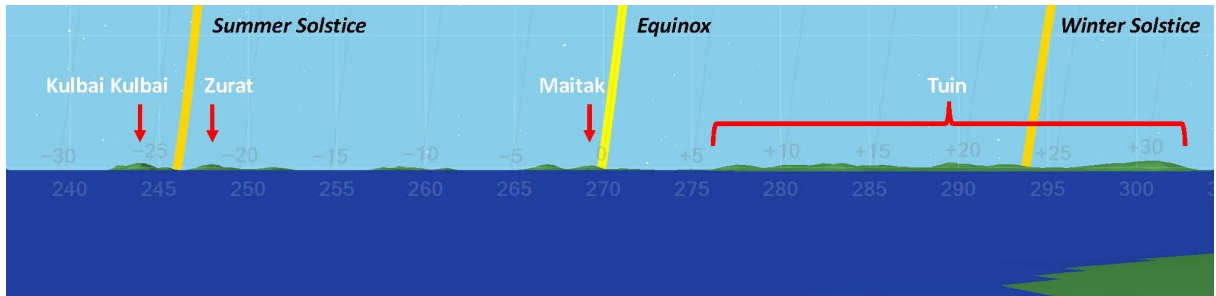

**Fig. 2:** A simulation of the setting Sun as seen from Kubin village, taken from the Horizon software package, developed by Andrew Smith at the University of Adelaide (www.agksmith.net/horizon/).

On Mer (Murray Island), Dauareb elder Segar Passi explained to Hamacher that he observes the sunset throughout the year from the front patio of his home. Rather than observing natural objects in the distance, he uses nearby reference points, such as light poles, houses, and trees. Passi explained how he predicts seasonal change using these reference points, demonstrating the adaptive nature of these observations to suit personal situations while still maintaining traditional meaning.

**2.2   Case Study 2: Solar Observations in Northern New South Wales**

The second author (Fuller) has been conducting ethnographic fieldwork with Saltwater Aboriginal people of coastal New South Wales (NSW) with reference to astronomical knowledge and Songlines for his doctoral research at the University of New South Wales (Fuller 2020). A few indicative observations of solar points on the North coast of NSW have emerged through this work. The first possibility was raised in a magazine article by Mary Gilmore (1932: 370):

> *"I well remember my father's astonishment and sense of discovery at finding that the natives knew the solstices just as truly as we did. I was too young to keep in mind all that he said about how they measured for the period, except that it was done by means of certain fixed mountain-rocks known to the tribes. One of these was somewhere near the head of the Clarence River. It was a rock mass that neither earthquake nor landslide could shake out of position. When the sun's edge at setting just touched the down-line on one side of this rock, it marked the period of the sun's turning […] The solstice was either just then, or within so many days of that. Watch was kept by those chosen for this duty, which was the utmost importance, the year being measured by it, and tribal ceremonies dependent on it for date."*

Gilmore, whose father was a cattle and sheep station manager, would have been living near Yamba, NSW on the Clarence River at the time. The account is clear that Aboriginal people were aware of the solar range and used the extreme points at the solstices to track time to inform ceremony. An important element of this account is the note that Aboriginal people based the year on the solstices, just as in Western society. No further details are provided and we must consider the context of this tiny fragment of knowledge, but it does provide some insight into divisions of time in that Aboriginal culture.

Archaeological site cards from the *Aboriginal Heritage Information Management System* (AHIMS) database for NSW were examined for locations of recorded sites that fit the description of "certain fixed mountain-rocks" in the area. This led to the possibility of a site at the head of the Clarence River, which is yet to be found. A non-AHIMS report from a heritage survey showed an arrow-shaped rock pointing to the West, and included a comment by the survey team that it pointed "towards Mt. Pikapeen", an isolated peak some 20 km to the West. It is unlikely that this is the site referred to by Gilmore's father, and it is suspected that the rock is pointing at Mt. Pikapeen (the peak to the left) and not a more local "rock mass" as described in Gilmore's account. It is worth noting that the full sunset range fits within the valley between Mt. Pikapeen and the Cambridge Plateau. Using the Horizon software, a simulation was made of the setting Sun from this site showing the Solstices and the Equinox (Fig. 3A; see Section 2.4).

Another potential observing point was suggested to Fuller by a cultural person from the Gumbayngirr community near Coffs Harbour who said that Mt. Coramba may have been used as a marker for the equinox sunrise as viewed from the high ground to the West of Coramba. Fuller visited the site and found no evidence of material culture present (such as a stone arrangement), but could clearly see the mountain with no sight impediments. Fig. 3B shows the view of the sunrise from this position. Further investigation is ongoing.

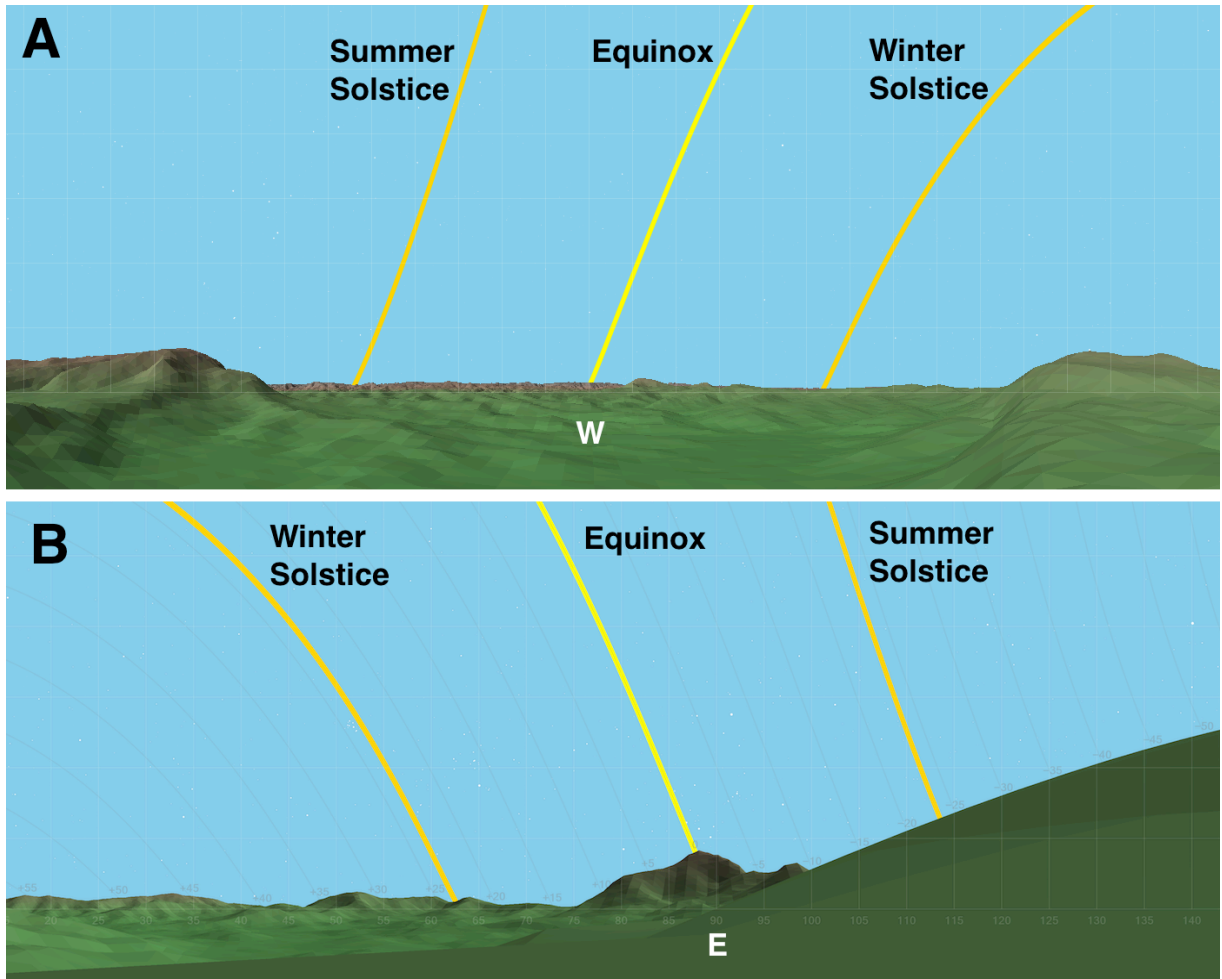

**Fig. 3:** (A) Horizon 3D profile from the arrow-shaped rock (West centred), with Mount Pikapeen in the northwest. (B) Equinoctial sunrise over Mt. Coramba in Gumbayngirr country.

### 2.3   Case Study 3: The Wurdi Youang Stone Arrangement, Victoria

The Wurdi Youang stone arrangement (hereafter WYSA, also known as the Mount Rothwell Archaeological Site) is located near the Little River, north of the You Yangs range between Melbourne and Geelong. This is the traditional land of the Wathaurong (Wadda-Warrung) people. The name "Wurdi Youang" means "big hill" (Smyth, 1878: 199), referring to the tallest of the You Yangs, and is known today as "Station Peak" or "Flinders Peak". In the 1800s, the region containing the You Yangs and the stone arrangement was called the Parish of Wurdi Youang in the County of Grant (Victoria Department of Crown Lands and Survey, 1928), from which the site draws its name. Aboriginal clan of the You Yangs area were the *Yawangi* and those in the area west of the Little River were called *Worinyaluk* (Massola 1969:11-13).

The site consists of approximately 100 basalt stones organised into an ovate arrangement, 50 m along its major axis, which itself is aligned East/West. The stones vary in height, with the tallest standing about 0.75 m high and individual stones weighing upwards of 500 kg (with an estimated total mass of 23 tonnes). The landscape is fairly barren, with bedrock sticking above the soil throughout the area. The terrain undulates and the highest point of the arrangement is near the western apex, which slopes downward to the northeast. It lies ~50 m from the Little River and is in a rain shadow with Mount Rothwell to the south and the Brisbane ranges to the west/southwest. The immediate surrounding area is farmland and the stones are relatively close to large power lines. The arrangement itself has been preserved and is protected (Lane and Fullagar 1980).

In August 1971, schoolteacher Louis N. Lane and a friend stumbled upon the stones (Lane, 1975). After years of study and lobbying, the arrangement (then called the Mount Rothwell Stone Arrangement) was granted heritage status in 1977 by the Victorian Archaeological Survey (AAV Site No. 7922-001), the first Aboriginal site in the state to achieve this classification. Lane had considered the site may contain astronomical alignments and wrote to prominent archaeoastronomer Alexander Thom in the UK in 1973 for his thoughts, enclosing photographs and survey details. These documents and records are now housed in the library archives of Deakin University in Geelong.

Wathaurong elders and custodians claim no or limited knowledge about the construction and use of the site, but do note it as a sacred place. The plot of land on which the arrangement is found was first owned (in colonial terms) by pastoralist and businessman Monckton Synnot (1826-1879) (Shaw, 1976; Victoria Department of Crown Lands and Survey, 1855). In the 1980s, local historian John Morieson began studying the site and hypothesised that outlier stones west of the primary arrangement aligned to the position of the setting Sun at the solstices and equinoxes (Morieson 2003). In 2006, Ray and Priscilla Norris worked closely with local custodian Trevor "Reg" Abrahams to survey the arrangement in detail and test this hypothesis.

Working with this paper's lead author (Hamacher), their survey and analysis concluded that Morieson's hypothesis was plausible and statistically supported. They also noted that the linear northerly and southerly sides of the arrangement aligned to the solstices if observed from its eastern apex, and the Sun sets at equinox directly over the middle of three prominent, 0.6 m high stones at the western apex (which is also the highest point of the arrangement; Fig. 4). Norris *et al.* (2013) published the study in *Rock Art Research*, confirming Morieson's hypothesis and showing that the alignments are accurate to within a few degrees.

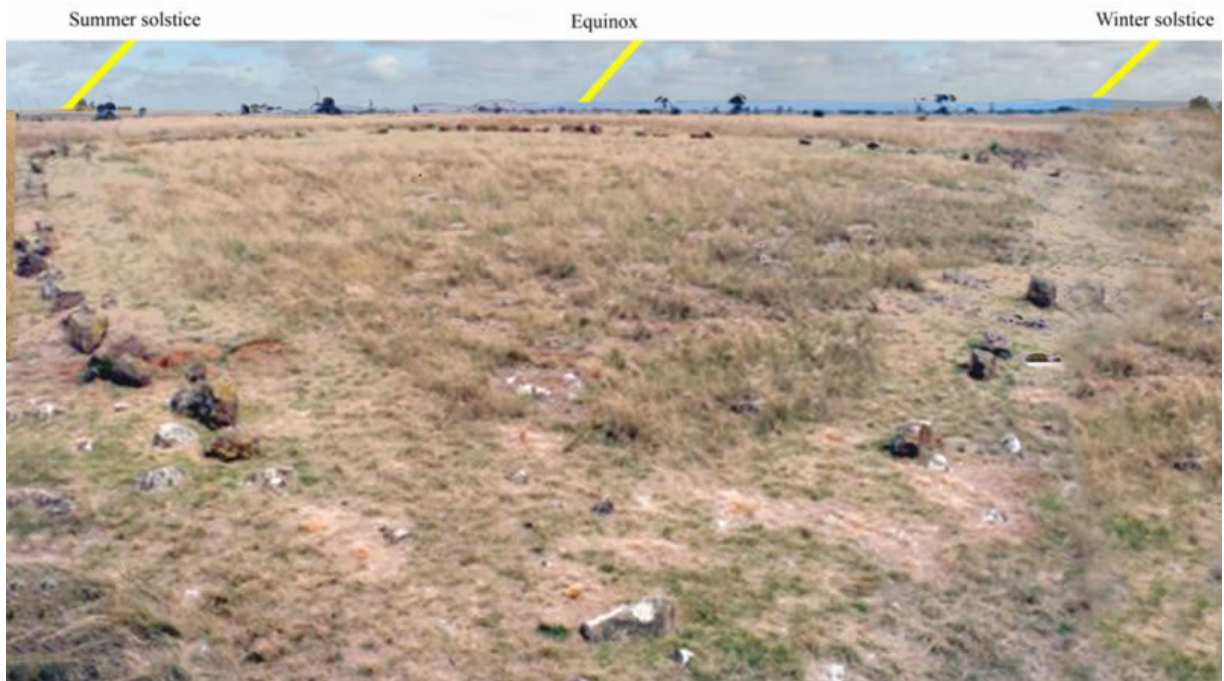

**Fig. 4.** The Wurdi Youang stone arrangement and the hypothesised solar alignments, from Norris et al. (2013). The point where the lines converges is the best vantage point for observing the solar markers.

The authors utilised a Monte Carlo statistical approach, simulating 10,000 alignments within the arrangement and noting their frequency. Results show that the alignments were probably constructed deliberately, with the simulations showing a 0.25% probability they are the result of chance. While this does not *definitively* prove the WYSA was constructed to mark the solar points, it *supports* the hypothesis that it was, at least in-part. In the absence of any traditional knowledge about the site and its meaning passed down to current Traditional Owners through oral tradition, the solar-hypothesis remains uncertain.

### 2.4 Case Study 4: Horizon Profiles in Wiradjuri Country, New South Wales

The third author (Leaman) has been conducting fieldwork with Wiradjuri (var: Wiradyuri) communities in the Central West region of New South Wales since 2014 to learn about their astronomical knowledge for his doctoral dissertation at the University of New South Wales. Part of this research has been to investigate the relationship between the land and sky with respect to cultural sites. One aspect of this has been to test the hypothesis that some cultural sites were selected geographically so that the "horizon profile" could serve as a marker for the solstices and equinoxes. Little research has been published suggesting that Aboriginal cultural sites were used with an astronomical function in mind, so testing this hypothesis utilises a newly-developing statistical approach.

Previous studies (e.g. Hamacher *et al.* 2012, Fuller *et al.* 2013, Norris *et al.* 2013) employed Monte Carlo statistics to argue that some stone arrangements and ceremonial grounds throughout New South Wales and far southeast Queensland were constructed with cardinal and astronomical alignments in mind. This includes sites in Wiradjuri country. However, the fragmentary nature of the ethnoarchaeological record, coupled with the damage or loss of cultural knowledge through the effects of invasion, colonisation, and forced displacement (e.g. see MacDonald 1998; Read 1983; 1984) hampers efforts to confirm this.

To test this hypothesis, Leaman and Hamacher (2018) developed a methodology to test the hypothesis that cultural sites of a particular type were constructed at a location where the horizon profile could mark the solstice and/or equinox points at dusk or dawn. The original intent was to produce a database of sites having a higher chance of astronomical alignments. The purpose was to select targeted sites for costly fieldwork studies. This methodology works on the assumption that, statistically, cultural sites used for astronomical observations are more likely to be placed within landscapes that take advantage of 'notch' and 'point' alignments of the Sun as it rises and/or sets throughout the year with respect to the local horizon (Fig. 5). The methodology utilises the *HORIZON* archaeoastronomical software[1] to generate horizon profiles centred on the cultural site being investigated.

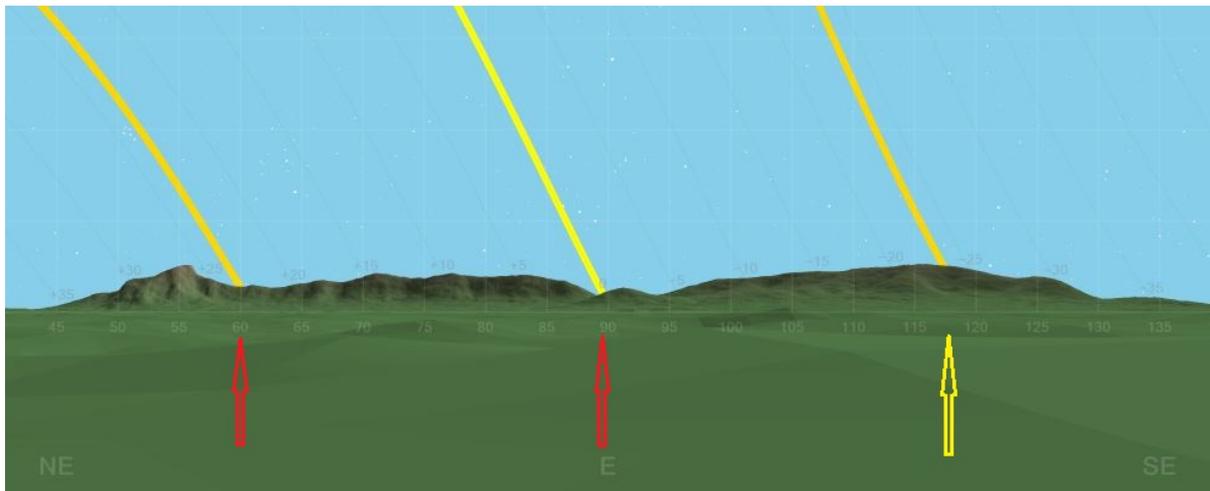

**Fig. 5:** Eastern horizon profile of the Kengal Lake Gilman site showing solar alignments with **"Notches"** (red arrows) and "Points" (yellow arrow). Yellow lines indicate paths of the Sun at the June (left), December (right) Solstice and Equinox (centre).

Cultural sites are ranked on a simple 10-point scale based on the number of potential horizon alignments with the solstices and equinoxes centred on notches and peaks, calculated using the following formula:

---

[1] www.agksmith.net/horizon/

$$R_S = (A_n/A_T) \times 10 \text{ (rounded to nearest whole number)}$$

where $R_S$ is the solar site rank, $A_n$ is the number of observed alignments (±0.5 deg), and $A_T$ is the total number of possible alignments. For example, for a hypothetical site with alignments to the Sun rising from a peak or notch on the December and June solstice, and a setting at a peak or notch on December solstice and the equinox, the ranking for this site would be:

$$R_s = (4/6) \times 10 = 6.67 \rightarrow 7$$

Leaman & Hamacher (2018) conducted a preliminary study of 14 Wiradjuri cultural sites, showing that ceremonial sites (particularly those associated with stone arrangements) generally have a higher ranking than sites with a more utilitarian function, such as stone quarries (Fig. 6). As a test to determine if this ranking was the result of chance, four randomly selected alternative sites using a random number generator were tested that fell within a 10 km² box centred on each cultural site. An analysis with a larger dataset of known and randomly selected sites is required for a more rigorous study, but preliminary results indicate a higher probability that sites with high rankings were deliberately chosen to take advantage of the horizon profile for solar observations (Fig. 7).

Leaman and Hamacher stress that the results are indicative of probability only, and that a low rank (i.e. where only one solar alignment occurs or is observed) cannot alone rule out that such observations were indeed made at the site. The methodology is considered complimentary to other methods, such as archival studies, ethnography, and on-site surveys. But it does suggest that Aboriginal people may have taken advantage of the sunrise and set points along the horizon when choosing cultural sites. The specific reasons for this, however, are not yet known.

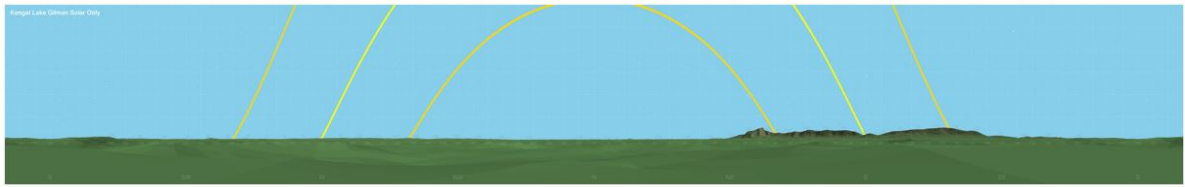

Kengal Lake Gilman (Solar Site Rank = 8; Solar Alignments: East =3; West = 2)

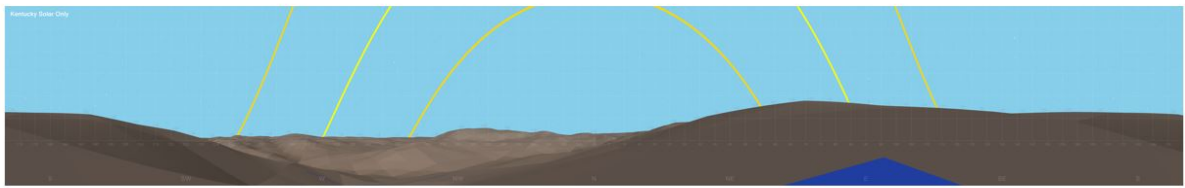

Kentucky Stone Arrangement (Solar Site Rank = 5; Solar Alignments: East = 1; West = 2)

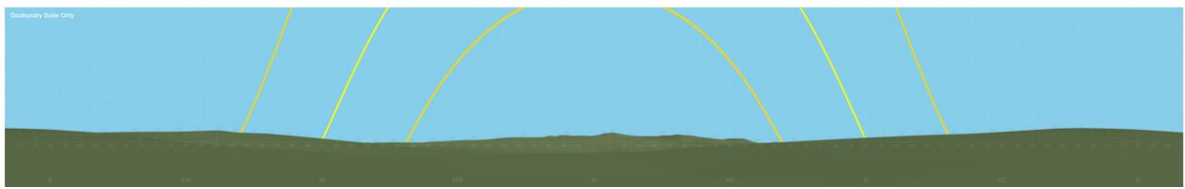

Goobundry Quarry (Solar Site Rank = 2; Solar Alignments: East = 1; West = 0)

**Fig. 6.** Typical horizon profiles used in the analysis, showing cultural sites of high (Kengal Lake Gilman), medium (Kentucky Stone Arrangement) and low (Goobundry Quarry) likelihood of being used for solar observations. The images are centred at due North.

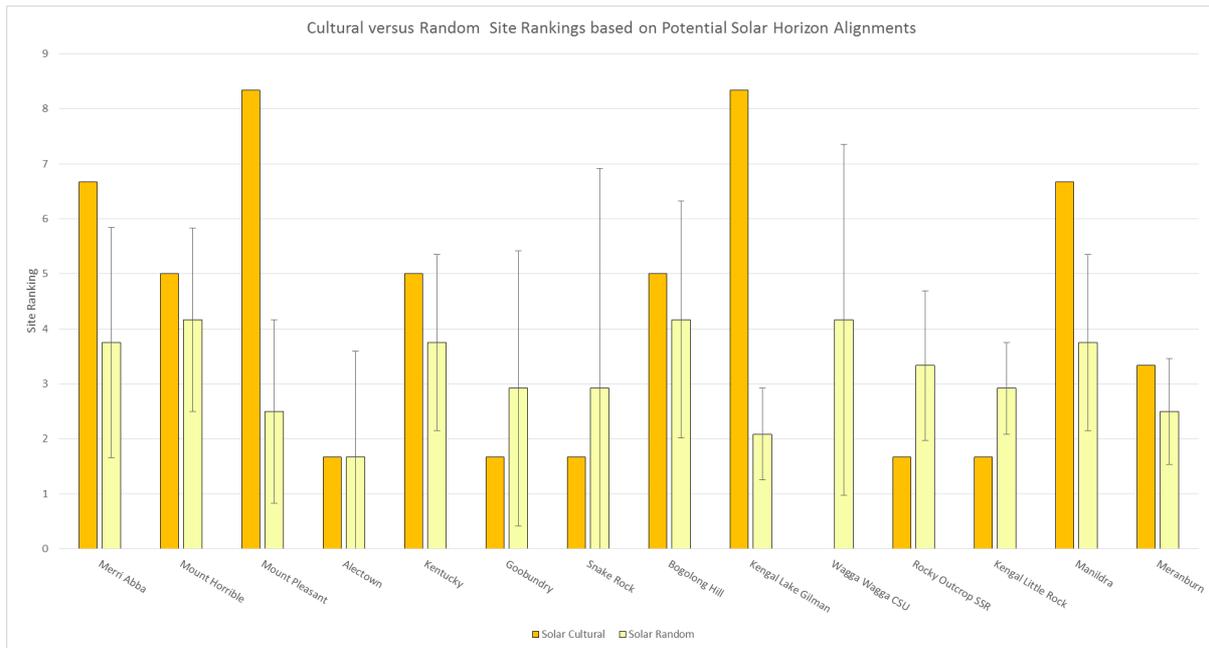

**Fig. 7:** Solar site rankings for 14 cultural sites studied. Error bars on random Sites represent the standard deviation of site ranking over four randomly selected alternative sites within 10 km² of the cultural sites. (Modified from Leaman and Hamacher, 2018).

## 3    Discussion and Conclusion

We see from evidence collected from ethnographic, historic, and archaeological studies that Aboriginal and Torres Strait Islander people observed and noted the position of the rising and/or setting position of the Sun throughout the year, and utilised that for practical purposes, such as tracking seasons. This was done using both human-made structures and the natural horizon landscape.

With the exception of Mua Island in the Torres Strait, details about the exact ways solar positions are tracked and the use of this information is scant and unclear. The Mua case study is the only definitive evidence that explains both how and why Indigenous Australians observe solar rise/set points throughout the year. To date, the study of the Wurdi Youang stone arrangement and sites in Wiradjuri country represent the only significant research outputs regarding Aboriginal observations of solar points, but they remain working hypotheses based on statistical analysis.

By examining ethnographic and historic information, we can definitively say that Aboriginal and Torres Strait Islander people do observe and note the positions of the rising and setting Sun throughout the year, with special reference to the solstices. Understanding how and why in more depth are topics of ongoing research.

**Acknowledgements**

The authors wish to pay respect to Aboriginal and Torres Strait Islander elders, past and present, and fully recognise and honour their intellectual property and traditions. They would like to thank (alphabetically) Trevor "Reg" Abrahams, Marcus Ferguson, Chris Heckenberg, Greg Ingram, James Ingram Jr., Mer Gemked Li, Segar Passi, David Towney, Larry Towney, and Peter White for comments, support, and input.

Leaman and Fuller are completing PhD programs at the University of New South Wales under the supervision of Duane Hamacher and Daniel Robinson. Fuller receives funding from the Australian Government Research Training Program Scholarship and ethnographic fieldwork was approved under UNSW Human Research Ethics project HC16335. Leaman receives funding from the Australian Government Research Training Program Scholarship and Central Tablelands Local Lands Services project CT00156, and ethnographic fieldwork was approved under UNSW Human Research Ethics project HC15037. Cr Bosun is an artist on Mua working with Hamacher on the Australian Research

## About the Authors

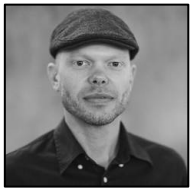
Dr Duane Hamacher is Associate Professor of Cultural Astronomy in the ASTRO-3D Centre of Excellence in the School of Physics at the University of Melbourne.

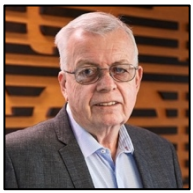
Robert Fuller is a PhD candidate at the University of New South Wales. His thesis subject is The Astronomy and Songline Connections of the Saltwater Aboriginal Peoples of the New South Wales Coast.

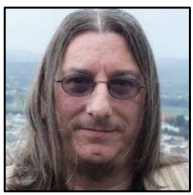
Trevor Leaman is a PhD candidate at the University of New South Wales. His thesis subject is Wiradjuri Astronomical Knowledge from central New South Wales.

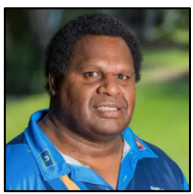
Councillor David Bosun is a culture man and artist from Mua in the western Torres Strait. He earned degrees and diplomas from James Cook University and Cairns TAFE.